\documentclass{article}
\usepackage{aaai18}
\usepackage{times} 
\usepackage{helvet}
\usepackage{courier}
\usepackage{latexsym}
\usepackage{amsfonts,amssymb}
\usepackage{amsmath}
\usepackage{graphicx}
\usepackage{float}
\usepackage{array}
\usepackage{multirow}
\usepackage{url}

\title{Multi-Perspective Neural Architecture for Recommendation System}

\author{
	Han Xiao$^*$, Yidong Chen, Xiaodong Shi \\
	Department of Intelligence Science and Technology, Xiamen University, \\
	Fujian Province, 361005, PR China, \\
	$^*$ corresponding author, email: bookman@xmu.edu.cn
}

\date{}

\begin{document}

\maketitle

\begin{abstract}
Currently, there starts a research trend to leverage neural architecture for recommendation systems. Though several deep recommender models are proposed, most methods are too simple to characterize users' complex preference. In this paper, for a fine-grain analysis, users' ratings are explained from multiple perspectives, based on which, we propose our neural architecture. Specifically, our model employs several sequential stages to encode the user and item into hidden representations. In one stage, the user and item are represented from multiple perspectives and in each perspective, the representations of user and item put attentions to each other. Last, we metric the output representations of final stage to approach the users' rating. Extensive experiments demonstrate that our method achieves substantial improvements against baselines.
\end{abstract}


\section{Introduction}
In the era of information explosion, information overload is one of the dilemmas we are confronted with. Recommender systems (RSs) are instrumental to address this problem, because they assist the users to identify which information is more preferred \cite{Xue2017Deep}. Further, to achieve better modeling ability of users' preference, neural architectures that deep learning methods are employed \cite{He2017Neural, Xue2017Deep}. There emerge many latest researches in this trend, such as NeuMF \cite{He2017Neural} and DMF \cite{Xue2017Deep}. Basically, most methods represent the user and item in a hidden semantic manner and then metric the hidden representations to predict the rating by cosine similarity or Multi-layer Perceptron (MLP). 

Despite the success of previous methods, they are still too simple to characterize users' complex preference. For the example of movie recommendation, user usually considers the quality of a movie from multiple perspectives, such as acting quality and movie style. It means that all the perspectives make effects on the preference, which traditional neural methods are difficult to characterize. To tackle this problem, in this paper, we encode the user and item into hidden representations from multiple perspectives and then metric the hidden representations to predict the preference. 

However, there still exist two challenges for the encoding process: to model hierarchically organized perspectives and to capture the correlation between user and item.

First, the perspectives are hierarchically organized from specific elements to abstract summarization. For the example of movie domain, there are basic aspects such as actor, director and shooting technique, based on which, abstract aspects such as acting quality and movie style are constructed. In detail, movie style is decided by director and shooting technique, while actor and director mostly determine the acting quality. Regarding the neural model, the output of each perspective indicates the representations of user/item metric in that perspective. For example, the encoded representation of user in actor perspective represents the user's preference for actors, while the encoded representation of item in movie style perspective indicates the style of this movie. The representation in low-level should support the analysis in high-level, which motivates us to employ a hierarchical deep neural architecture.  Thus, it is reasonable to apply multiple sequential stages and to encode the user/item from multiple perspectives in each stage.

\begin{figure}
	\centering
	\includegraphics[width=\linewidth]{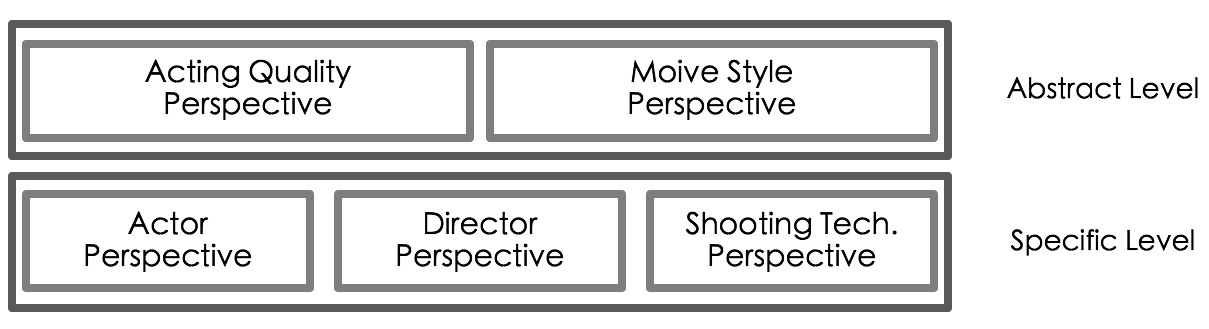}
	\caption{Hierarchically Organized Perspectives. The specific level representations support the analysis of abstract level representation. }
	\label{fig:hier}
\end{figure}

Second, the correlation between user and item is weak in the encoding process of current models. However, in fact, from the study of psychology \cite{carlson2009psychology}, users' preference is subjective and would be slightly adjusted according to a specific item, while the subjective feature of a specific item could be slightly different from different users' insight. Therefore, we employ the attention mechanism \cite{Schmidhuber2015Deep} to address the correlated effects between user and item.

Specifically in this paper, to model user's complex preference on item, we propose a novel neural architecture for top-N recommendation task. Overall, our model encodes the user and item into hidden semantic representations and then metrics the hidden representations into predicted preference degree with cosine similarity. Specifically, regarding the encoding process, our model leverages several sequential stages to model the hierarchically organized perspectives. In each stage, there exist several perspectives and in each perspective, the representations for user and item would adjust each other by attention mechanism. Besides, we have studied two methods for constructing the attention signal, which are listed as ``Softmax-ATT'' and ``Correlated-ATT''.

We evaluate the effectiveness of our neural architecture for the top-N recommendation task in six datasets from five domains (i.e. Movie, Book, Music, Baby Product, Office Product). Experimental results on these datasets demonstrate our model consistently outperforms the other baselines with remarkable improvements and achieves the state-of-the-art performance among deep recommendation models.

In summary, our contributions are outlined as follows:
\begin{itemize}
	\item We propose a novel neural architecture for recommendation systems, which focuses on the hierarchically organized perspectives and the correlation between user and item.
	\item To our best knowledge, this is the first paper to introduce attention mechanism into neural recommendation systems.
	\item Experimental results show the effectiveness of our proposed architecture, which outperforms other state-of-the-art methods in the top-N recommendation task.
\end{itemize}

The organization of this paper is as follows. First, problem formulation and related work are introduced. Second, our neural architecture is discussed. Third, we conduct the experiments to verify our model. Last, concluding remarks are in the final section.

\section{Problem Formulation \& Related Work}
Suppose there are $M$ users $\mathcal{U}=\{u_1, ..., u_M\}$ and $N$ items $\mathcal{V}=\{v_1, ..., v_N\}$. Let $R \in \mathbb{R}^{M \times N}$ indicate the rating matrix, where $R_{ij}$ is the rating of user $i$ on item $j$ and we denote $unk$ if it is unknown. There are two manners to construct the user-item interaction matrix $T \in \mathbb{R}^{M \times N}$, which indicates the user $i$ whether performs operation on item $j$ as 
\begin{eqnarray}
T_{ij} = 
\begin{cases}
0, & if~R_{ij}~is~unk \\
1, & otherwise
\end{cases}\label{eq1} \\
T_{ij} = 
\begin{cases}
0, & if~R_{ij}~is~unk \\
R_{ij}, & otherwise
\end{cases}\label{eq2}
\end{eqnarray}
Most traditional models for recommendation system employ Equation (\ref{eq1}) as the input to their models, \cite{Wu2016Collaborative, He2017Neural}, while some latest work takes the known entry as the ratings $R_{ij}$ rather than $1$ as Equation (\ref{eq2}) shows \cite{Xue2017Deep}. We apply the second  setting, because we suppose the explicit ratings in Equation (\ref{eq2}) could reflect the preference level of a user for an item.

The recommendation systems are conventionally formulated as the problem of estimating the rating of each unobserved entry in $Y$, which is leveraged to rank the items. Model-based approaches that are the mainstream methodology leverage an underlying model to generate all the ratings:
\begin{eqnarray}
	\hat{T}_{ij} = \mathcal{M}(u_i, v_j | \Theta)
\end{eqnarray}
where $\hat{T}_{ij}$ denotes the predicted score of interaction $T_{ij}$ between user $u_i$ and item $v_j$, $\Theta$ indicates the model parameter and $\mathcal{M}$ denotes the recommendation model that predicts the scores. With the predicted scores by model $\mathcal{M}$, we could rank the items for an individual user to conduct personalized recommendation. 

First, matrix factorization as semantic latent space methodology is proposed for this task. For the classical method of latent factor model \cite{Koren2009Matrix}, which basically applies the inner product of the hidden representations of user and item to predict the entity $\hat{T}_{ij}$ as follows
\begin{eqnarray}
\hat{T}_{ij} = \mathcal{M}_{LFM}(u_i, v_j | \Theta) = p_i^T q_j \label{eq3}
\end{eqnarray}
where $\hat{T}$ means the predicted score, $\mathcal{M}_{LFM}$ indicates latent factor model, $p_i$/$q_j$ indicates the hidden representation of user $u_i$ / item $v_j$, respectively.
Also, there follow many related researches such as \cite{Koren2008Factorization, Mcauley2013Hidden, Bao2014TopicMF}.

Then, extra corpus such as social relationship is incorporated into recommendation for a further improvement, \cite{Ma2008SoRec}. However, because the additional corpus is difficult to obtain and is often full of noise, this methodology is still under limitation.

Last, due to the powerful representation learning ability of neural network, deep learning methods have been successfully applied into this field. Restricted Boltzmann Machines \cite{Salakhutdinov2007Restricted} are the pioneer for this branch. Meanwhile, autoencoders and the denoising autoencoders have also been investigated for this task, \cite{Li2015Deep, Sedhain2015AutoRec, Strub2015Collaborative}. The main principle of these methods is to predict user's ratings through learning hidden representations with historical behaviors (i.e. ratings and reviews). 

Recently, to learn non-linear interactions, neural collaborative filtering (NeuCF) \cite{He2017Neural} presents an approach, where users and items are embedded into numerical vectors and then the embeddings are processed by a multi-layer perceptron to learn the users' preference. Deep matrix factorization (DMF) \cite{Xue2017Deep} jointly takes the spirit of latent factor model and neural collaborative filtering method. Specifically, DMF independently encodes the user and item by multi-layer perceptron (MLP) and then metrics the hidden representations of user and item from the MLP in the manner of Equation (\ref{eq3}) to predict the preference degree. In fact, DMF takes the advantage of deep representation learning to achieve the state-of-the-art performance.

There list the notations used in the following sections. $u$ indicates a user and $v$ indicates an item. $i$ and $j$ are the index for $u$ and $v$, respectively. $T$ denotes the user-item interaction matrix, formulated in Equation (\ref{eq2}), while $T^+$ denotes the observed interactions, $T^-$ means all zero elements in $T$ and $T^-_{sample}$ denotes the negative instances generated from sampling. Notably, $T^+ \bigcup T^-_{sample}$ means the training and developing dataset while $T^-$ is the source of testing dataset. Further, we indicate the $i$-th row of matrix $T$ as $T_{i*}$, $j$-th column as $T_{*j}$ and its $(i,j)$-th entry as $T_{ij}$.

\section{Methodology} 
In this section, first, we will introduce the overall sketch of our proposed neural architecture, which is illustrated in Fig.\ref{fig:overall}. Then, we will discuss the details of each component in a bottom-up manner, namely interaction matrix, sequential stages and cosine similarity. Also the implementation of each stage and attention mechanism (demonstrated in Fig.\ref{fig:attention1} and Fig.\ref{fig:attention2}) will be analyzed as follows. Last, we present our loss function and training algorithm.

\subsection{Neural Architecture}
\begin{figure}[H]
	\centering
	\includegraphics[width=0.95\linewidth]{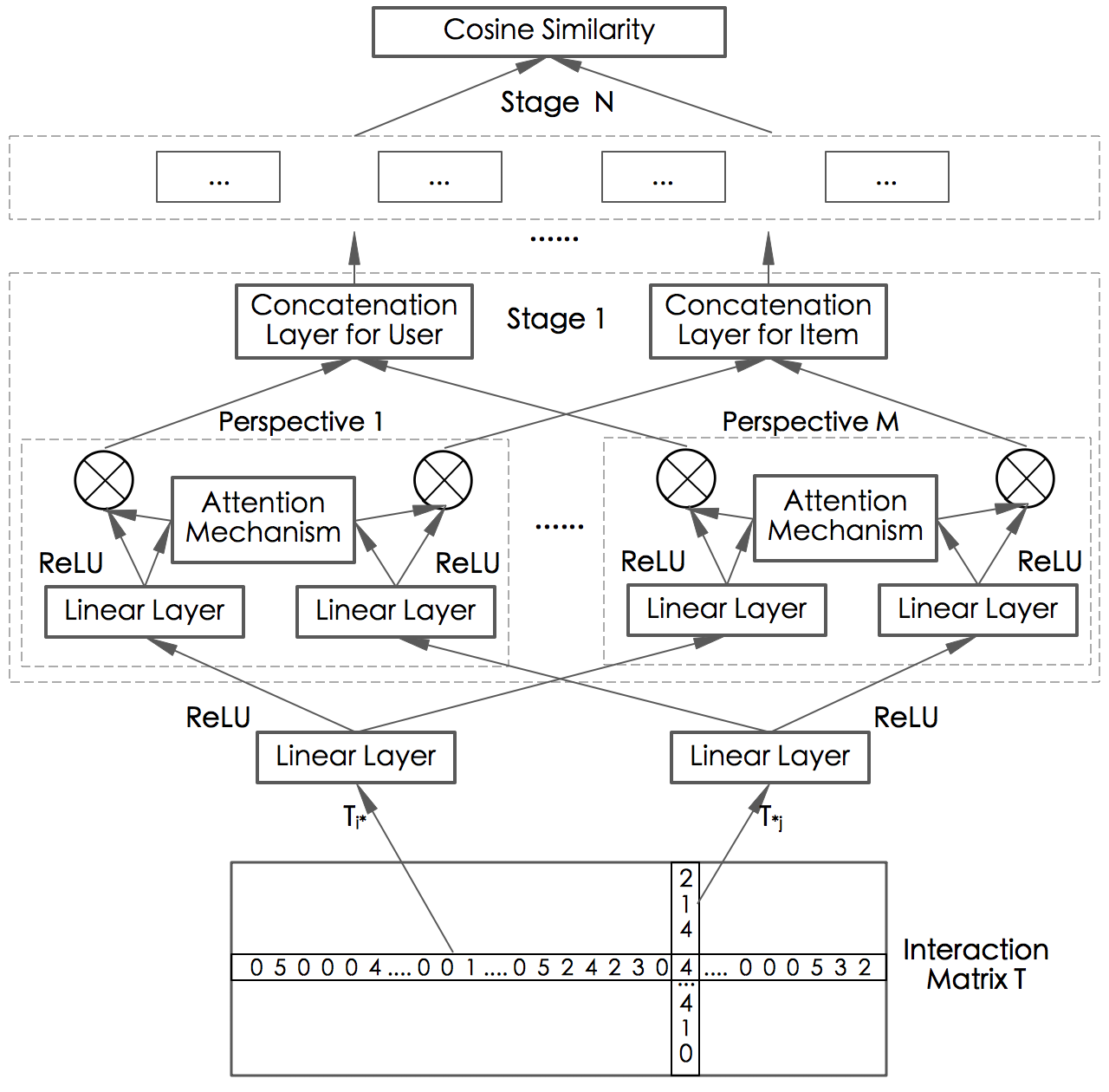}
	\caption{Proposed Neural Architecture. We leverage the corresponding row/column of interaction matrix as the input of user/item. To characterize hierarchically organized perspectives, we employ several sequential stages to encode the input. In each stage, there exist several perspectives. In each perspective, the input of this perspective will be encoded into hidden representations by linear transformation with ReLU activation function and then attention mechanism addresses the correlations for the encoded representations of user/item to generate the output of this perspective. Furthermore, the outputs of all the perspectives in one stage are respectively concatenated as the output representation of user and item for this stage. Finally, the representation of user and item would be metric by cosine similarity to predict the user's preference.}
	\label{fig:overall}
\end{figure}
Our neural architecture is demonstrated in Fig.\ref{fig:overall}. Basically, our model is composed by three components, namely interaction matrix, sequential stages and cosine similarity.

\textbf{Interaction Matrix.} Mentioned in previous section, we form the interaction matrix as Equation (\ref{eq2}), which is the input of our model. From the interaction matrix $T$, each user $u_i$ is represented as a high-dimensional vector $T_{i*}$, which indicates the corresponding user's ratings across all items, while each item  $v_j$ is represented as a high-dimensional vector $T_{*j}$, which means the corresponding item's ratings across all users. Notably, it is a conventional trick to fill the unknown entry as $0$. To overcome the sparsity of interaction matrix, the inputs of user and item are transformed by linear layer with the activation function ReLU (i.e $f(x)= max(x, 0)$) as 
\begin{eqnarray}
	\mathbf{r_u} = \sigma (\mathbf{WT_{i*} + b_u}) \\
	\mathbf{r_v} = \sigma (\mathbf{MT_{*j}^\top + b_v}) 
\end{eqnarray}
where $\mathbf{r_u/r_v}$ is the output of this layer for user/item, $\mathbf{T}_{i*}$/$\mathbf{T}_{*j}$ means the input of row/column-specific interaction matrix for user/item, $\mathbf{W, M, b_u, b_v}$ are the parameters of linear layer and $\sigma$ is the activation function (i.e. ReLU).

\textbf{Sequential Stages.} In order to model the hierarchically organized perspectives shown in Fig.\ref{fig:hier}, we leverage multiple sequential stages, shown in Fig.\ref{fig:overall}. In each stage, there exist several perspectives to model the user/item representations from multiple aspects. In each perspective, the output of last stage is regarded as the input of this perspective while the outputs of all the perspectives in one stage are respectively concatenated as the output representation of user and item for this stage, shown in Fig.\ref{fig:overall}.

Specifically in one perspective, first, the inputs of this perspective that the output representations of user and item in last stage are transformed by linear layer with the activation function \textit{ReLU}. 
\begin{eqnarray}
\mathbf{q_{u,s,p}} =  \sigma (\mathbf{W_{s,p} r_{u,s-1} + b_{u,s,p}})  \\
\mathbf{q_{v,s,p}} =  \sigma (\mathbf{M_{s,p} r_{v,s-1} + b_{v,s,p}})  
\end{eqnarray}
where $\sigma$ indicates the ReLU function, $\mathbf{q_{u,s,p}}$/$\mathbf{q_{v,s,p}}$ is the output for user/item of linear layer in $p$-th perspective of $s$-th stage, $\mathbf{r_{u,s-1}}$/$\mathbf{r_{v,s-1}}$ is the output for user/item of last stage and $\mathbf{W_{s,p}, M_{s,p}, b_{u,s,p}, b_{v,s,p}}$ are model parameters.

Then, attention signal is generated from the output of linear layer by attention mechanism. 
\begin{eqnarray}
\mathbf{a_{u,s,p}} = \mathcal{A}_u(\mathbf{q_{u,s,p}, q_{v,s,p}}) \label{eq4} \\
\mathbf{a_{v,s,p}} = \mathcal{A}_v(\mathbf{q_{u,s,p}, q_{v,s,p}}) \label{eq5}
\end{eqnarray}
where $\mathbf{a_{u,s,p}}$/$\mathbf{a_{v,s,p}}$ is the attention signal for user/item in $p$-th perspective of $s$-th stage and $\mathbf{q_{u,s,p}}$/$\mathbf{q_{v,s,p}}$ is the output for user/item of linear layer in $p$-th perspective of $s$-th stage. $\mathcal{A}_u$/$\mathcal{A}_v$ indicates the attention function for user/item.

Last, the output of this perspective is generated by weighting the output of linear layer with the attention signal in the manner of element-wise product. Mathematically, we have:
\begin{eqnarray}
\mathbf{r_{u,s,p}} = \mathbf{q_{u,s,p}} \otimes \mathbf{a_{u,s,p}} \\
\mathbf{r_{v,s,p}} = \mathbf{q_{v,s,p}} \otimes \mathbf{a_{v,s,p}}
\end{eqnarray}
where $\mathbf{r_{u,s,p}}$/$\mathbf{r_{v,s,p}}$ is the output of the $p$-th perspective in $s$-th stage, $\mathbf{a_{u,s,p}}$/$\mathbf{a_{v,s,p}}$ is the attention signal for user/item in $p$-th perspective of $s$-th stage and $\mathbf{q_{u,s,p}}$/$\mathbf{q_{v,s,p}}$ is the output for user/item of linear layer in $p$-th perspective of $s$-th stage. $\otimes$ means the element-wise product.

\textbf{Cosine Similarity.} To generate the user's $u_i$ preference on  the item $v_j$, we measure the output representations of user/item in the final stage with cosine similarity, which is a conventional operation in neural architecture, \cite{Wang2016Semi}, mathematically as 
\begin{eqnarray}
	\hat{T}_{ij} & = & cosine(\mathbf{r_{u,final}, r_{v,final}}) \nonumber \\
	& = & \frac{\mathbf{r_{u,final}}^\top\mathbf{r_{v,final}}}{||\mathbf{r_{u,final}}||~||\mathbf{r_{v,final}}||}
\end{eqnarray}
where $\hat{T}_{ij}$ is the predicted preference of user $u_i$ on item $v_j$, $\mathbf{r_{u,final}}$/$\mathbf{r_{v,final}}$ is the output representation for user/item of the final stage, $||\cdot||$ is the length of vector.

\subsection{Attention Mechanism} 
Motivated in Introduction, to characterize the correlations between user and item, we leverage attention mechanism to refine the encoded representations of user/item as Equation (\ref{eq4}) and Equation (\ref{eq5}) show. With the attention mechanism, the final representations for user/item are more flexible and more precise to characterize the user's complex preference on the item.

\begin{figure}[H]
	\centering
	\includegraphics[width=0.55\linewidth]{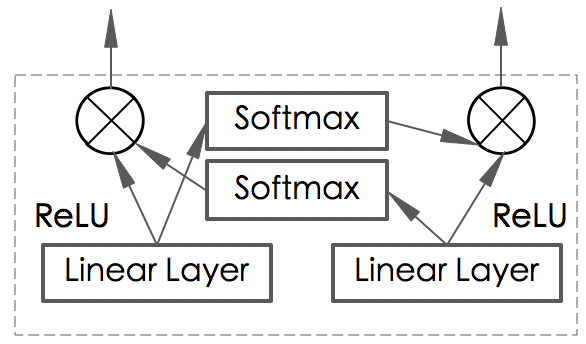}
	\caption{The first attention mechanism that the ``Softmax-ATT'', which leverages a simple softmax layer to construct the attention signal.}
	\label{fig:attention1}
\end{figure}

Firstly, shown in Fig.\ref{fig:attention1}, we directly employ a softmax layer to construct the attention signal, which is a conventional and common form for attention-based methods, \cite{yang2017semi, cui2016attention, yin2015abcnn}, mathematically as:
\begin{eqnarray}
	\mathcal{A}_u(\mathbf{q_{u,s,p}, q_{v,s,p}})  = softmax(\mathbf{A_{u,s,p}q_{v,s,p}}) \\
	\mathcal{A}_v(\mathbf{q_{u,s,p}, q_{v,s,p}})  = softmax(\mathbf{A_{v,s,p}q_{u,s,p}})
\end{eqnarray}
where $\mathbf{A_{u,s,p}}$/$\mathbf{A_{v,s,p}}$ is the attention matrix for user/item in the $p$-th perspective of $s$-th stage, $softmax$ is the softmax operation for vector and other symbols are introduced in last subsection as $\mathcal{A}_u$/$\mathcal{A}_v$ is the attention function for user/item and $\mathbf{q_{u,s,p}}$/$\mathbf{q_{v,s,p}}$ is the output for user/item of linear layer in $p$-th perspective of $s$-th stage. 

Notably, the attention matrices are model parameters to learn. Specifically, the attention signal for user is generated from the representation of item, while the attention signal for item is generated from the representation of user, which accords to our motivation of correlation. We call this attention setting as ``Softmax-ATT''.

\begin{figure}[H]
	\centering
	\includegraphics[width=0.60\linewidth]{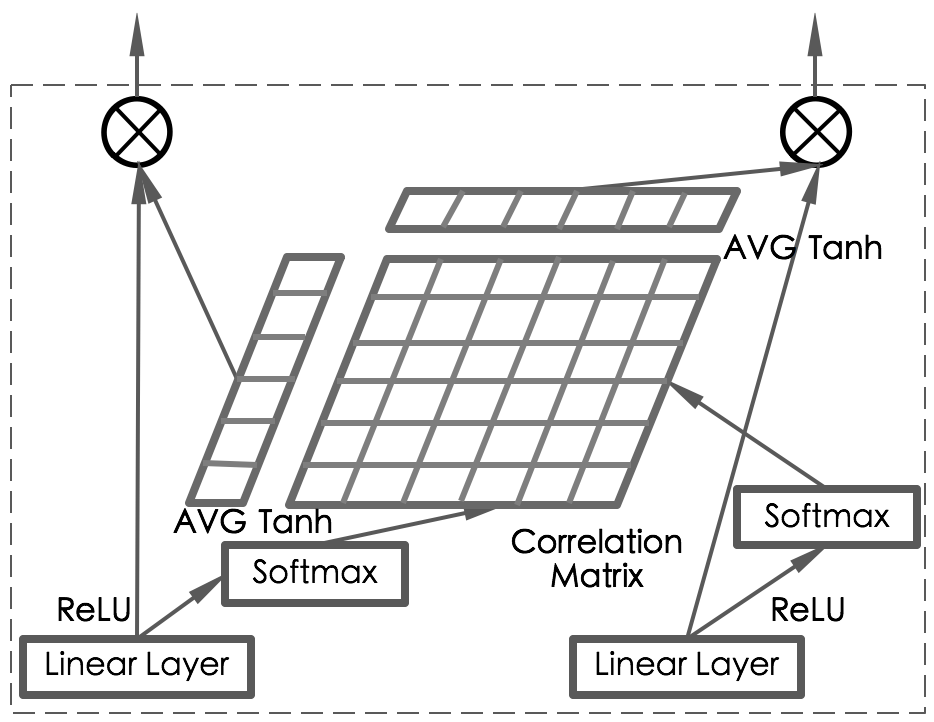}
	\caption{The second attention mechanism that the ``Correlated-ATT'', which leverages the correlation matrix to strengthen the correlation characterization.}
	\label{fig:attention2}
\end{figure}

However, the correlation modeled by simple softmax operation could still be improved. For a more effective correlation modeling, we propose a novel attention structure, shown in Fig.\ref{fig:attention2}. First, we compute the softmax vectors as the first attention method does:
\begin{eqnarray}
	\mathbf{a_{u,s,p}}  = softmax(\mathbf{A_{u,s,p}q_{v,s,p}}) \\
	\mathbf{a_{v,s,p}}  = softmax(\mathbf{A_{v,s,p}q_{u,s,p}})
\end{eqnarray}
where $\mathbf{a_{u,s,p}}$/$\mathbf{a_{v,s,p}}$ is the output of softmax layer in $p$-th perspective of $s$-th stage and other symbols are introduced previously. Then, we construct the correlation matrix between the representation of user and item, as 
\begin{eqnarray}
	\mathbf{C_{s,p}} =  \mathbf{a_{u,s,p}a_{v,s,p}^\top}
\end{eqnarray}
where $\mathbf{a_{u,s,p}}$/$\mathbf{a_{v,s,p}}$ is the output of softmax layer, $\mathbf{C_{s,p}}$ is the correlation matrix of $p$-th perspective in $s$-th stage, which contains the correlated information of all the dimensions for user/item. Last, we process the correlation matrix with $tanh$ activation function and average the row/column as the attention vector for user/item, as 
\begin{eqnarray}
	\mathcal{A}_u(\mathbf{q_{u,s,p}, q_{v,s,p}})  = average_{row}(tanh(\mathbf{C_{s,p}})) \\
	\mathcal{A}_v(\mathbf{q_{u,s,p}, q_{v,s,p}})  = average_{column}(tanh(\mathbf{C_{s,p}}))
\end{eqnarray}
where $average_{row}$/$average_{column}$ indicates the average operation for row/column and other symbols are introduced previously. With the explicit computation of correlation matrix, the correlated effects between user and item could be characterized to a better extent. We call this attention setting as ``Correlated-ATT''.

\subsection{Training} 
The definition of objective function for model optimization is critical for recommendation models. Specifically, regarding our model, we take advantage of point-wise objective function and cross-entropy loss. Actually, though the square loss is largely performed in many existing models, \cite{hu2008collaborative, mnih2008probabilistic}, neural architectures usually employ cross-entropy loss \cite{he2017generating, wu2017sequence}. Thus, our objective function $\mathcal{L}$ is as 
\begin{eqnarray}
\mathcal{L} = \sum_{(i,j) \in T^+ \bigcup T^-} T_{ij} log \hat{T}_{ij} + (1 - T_{ij}) log (1 - \hat{T}_{ij})
\end{eqnarray}
where $\mathcal{L}$ is the objective function, $T$ is the golden rating, $\hat{T}$ is the predicted score and other symbols are introduced in Related Work.
Specifically as previous literatures \cite{he2017generating, Xue2017Deep}, the target value $T_{ij}$ is a binarized $1$ or $0$ for the rating $R_{ij}$, denoting whether the user $u_i$ has interacted with item $v_j$ or not. Besides, the model is trained using Stochastic Gradient Descent (SGD) with Adam \cite{kingma2014adam}, which is an adaptive learning rate algorithm.

The training process needs the negative samples and all the ratings in the training set are the positive ones. Thus, we randomly sample several negative samples that are not in the training/developing/testing dataset for one positive sample. Besides, we apply the concept of negative sample ratio to illustrate how many negative samples would be generated for one positive instance.

\begin{table*}
	\centering
	\renewcommand\arraystretch{1.1}
	\begin{tabular}{cc|cc|cc|cc}
		\hline
		\hline \multirow{2}{*}{Datasets} & \multirow{2}{*}{Metrics} & \multicolumn{2}{c|}{Baselines} & \multicolumn{2}{c|}{Our Methods} & \multicolumn{2}{c}{Improvements over the Best Baseline} \\
		\cline{3-8} & & NeuMF & DMF & Softmax-ATT & Correlated-ATT &  Softmax-ATT & Correlated ATT\\
		\hline 
		\hline \multirow{2}{*}{Movie} & NDCG & 0.395 & 0.400 & 0.402 & \textbf{0.410} & 0.50\% & 2.50\% \\
		& HR  & 0.670 & 0.676 & 0.686 & \textbf{0.688} & 1.48\% & 1.78\% \\
		\hline \multirow{2}{*}{Movie-1M} & NDCG & 0.440 & 0.445 & 0.447 & \textbf{0.448} & 0.45\% & 0.67\% \\
		& HR & 0.722 & 0.723 & 0.732 & \textbf{0.735} & 1.24\% & 1.66\% \\
		\hline \multirow{2}{*}{Book} & NDCG & 0.477 & 0.471 & 0.483 & \textbf{0.484} & 1.26\% & 1.47\% \\
		& HR & 0.676 & 0.667 & 0.690 & \textbf{0.694} & 2.07\% & 2.81\% \\
		\hline \multirow{2}{*}{Music} & NDCG & 0.220 & 0.230 & 0.253 & \textbf{0.262} & 10.00\% & 13.90\% \\
		& HR & 0.371 & 0.382 & 0.428 & \textbf{0.445} & 12.04\% & 16.49\% \\
		\hline \multirow{2}{*}{Baby} & NDCG & 0.160 & 0.162  & 0.172 & \textbf{0.182} & 6.17 & 12.34\% \% \\
		& HR & 0.285 & 0.287  & 0.321 & \textbf{0.366} & 11.85\% & 27.52\%  \\
		\hline \multirow{2}{*}{Office Product} & NDCG & 0.233 & 0.243 & 0.261 & \textbf{0.262} & 7.40\% & 7.81\% \\
		& HR & 0.518 & 0.520 & 0.521 & \textbf{0.532} & 0.19\% & 2.30\% \\
		\hline
		\hline
	\end{tabular}
	\caption{NDCG@10 and HR@10 Comparisons of Different Methods. We conduct t-test for statistical significance and $p<0.01$, which means all of the improvements are statistically significant.}
	\label{performance}
\end{table*}

\section{Experiment}
In this section, first, we will introduce the basic experimental settings, namely datasets, evaluation and implementation. Then, we will conduct the experiments about model performance. Last, we will analyze the sensitivity to hyper-parameters for our model.

\subsection{Experimental Setting} 
\textbf{Datasets.} We evaluate our models on six widely used datasets from five domains in recommender systems: MovieLens 100K (Movie), MovieLens 1M (Movie-1M), Amazon music (Music), Amazon Kindle books (Book), Amazon office product (Office) and Amazon baby product (Baby). \footnote{\url{https://grouplens.org/datasets/movielens/}} \footnote{\url{http://jmcauley.ucsd.edu/data/amazon/}} We process the datasets, according to the previous literatures \cite{Wu2016Collaborative, Xue2017Deep, He2017Neural}. For the datasets of Movie and Movie-1M, we do not process them, because they are already filtered. Besides, other datasets are filtered to be similar to MovieLens data: only those users with at least 20 interactions and items with at least 5 interactions are retained.\footnote{We will publish our filtered datasets, once accepted.} We list the statistics of all the six processed datasets in Tab.\ref{sta}.

\begin{table}[H]
	\centering
	\renewcommand\arraystretch{1.1}
	\begin{tabular}{c|c|c|c|c}
		\hline \textbf{Statistics} & \textbf{\#Users} & \textbf{\#Items} & \textbf{\#Ratings} & \textbf{Density} \\
		\hline 
		\hline Movie & 994 & 1.683 & 100,000 & 6.294\% \\
		\hline Movie-1M & 6,040 & 3,706 & 1,000,209 & 4.468\% \\
		\hline Music & 1,776 & 12,929 & 46,087 & 0.201\% \\
		\hline Book & 14,803 & 96,538 & 627,441 & 0.004\%\\
		\hline Office & 941 & 6,679 & 27,254 & 4.336\%\\
		\hline Baby & 1,100 & 8,539 & 30,166 & 0.321\%\\
		\hline
	\end{tabular}
	\caption{Statistics of Datasets.}
	\label{sta}
\end{table}

\textbf{Evaluation.} To verify the performance of our model for item recommendation, we adopted the \textit{leave-one-out} evaluation, which has been widely used in the related literatures \cite{He2017Neural, Xue2017Deep}. We held-out the latest interaction as the test item for each user and utilize the remaining dataset for training. Since it is too time-consuming to rank all the items for every user during testing, following \cite{Koren2009Matrix, He2017Neural, Xue2017Deep}, we randomly sample 100 items that are not interacted by the corresponding user as the test set for this user. Among the 100 items together with the test item, we get the rank according to the prediction scores. We also use \textit{Hit Ratio (HR)} and \textit{Normalized Discounted Cumulative Grain (NDCG)} to evaluate the ranking performance, \cite{Xue2017Deep, he2017generating}. As default, in our experiments, we truncate the rank list at 10 for both metrics, where HR/NDCG intuitively means HR@10/NDCG@10, as previous literatures \cite{Xue2017Deep}. It is the similar notation for HR@K/NDCG@K.

\textbf{Detailed Implementation.} 
We implement our proposed methods based on Tensorflow\footnote{\url{https://www.tensorflow.org}} and the released codes of DMF \cite{Xue2017Deep}. Our codes will be released publicly upon acceptance. To determine the hyper-parameters of our model, we randomly sample one interaction for each user as the developing data and tune hyper-parameters on it. For neural part of our model, we randomly initialize model parameters with a Gaussian distribution (with the mean of $0$ and standard deviation of $0.01$). 

We test the batch size of $[128, 256, 512, 1024]$, the negative instance number per positive instance of $[3, 7, 15]$, the learning rate of $[0.0001, 0.0005, 0.001, 0.005]$, the number of stage $[1,2,3,4]$, the number of perspectives in each stage $[4,6,8]$, the dimension of all the linear layers $[50, 100, 150]$, the dimension of the output of non-final stage $[50, 100, 150]$ and the dimension of the output of final stage $[8, 16, 32, 64, 128]$. The optimal settings for our model are listed as: batch size as $256$, negative instance number per positive instance as $7$, learning rate as $0.0001$, number of stage as $3$, number of perspectives of each stage as $6$, the dimension of all the linear layers as $50$, the dimension of the output of non-final stage as $50$ and the dimension of the output of final stage as $128$.

\subsection{Performance Verification} 
\textbf{Baselines.} As our proposed methods aim to model the relationship between users and items, we follow \cite{Xue2017Deep} and \cite{He2017Neural} to mainly compare with user-item models. Thus, we leave out the comparison with item-item models, such as CDAE \cite{Wu2016Collaborative}. Actually, since the neural recommendation methodology just starts to be focused, we just list two suitable latest baseline models.

\textbf{NeuMF.} This is a neural matrix factorization method for item recommendation. This method embeds the user and item as hidden representations and then leverages a multiple layer perceptron to learn the user-item action function based on the embeddings of user and item. We implement the pre-training version of NeuMF and tune its hyper-parameters in the same way as \cite{He2017Neural}.

\textbf{DMF.} This is the state-of-the-art neural recommendation method. This method encodes the user and item into hidden representations independently and metrics the representations between user and item to predict the user's preference degree for the item. We implement DMF and tune its hyper-parameters in the same way as \cite{Xue2017Deep}.

\textbf{Conclusions.} The comparisons are illustrated in Tab.\ref{performance}. Thus, we have concluded as below:
\begin{itemize}
	\item Our method outperforms the baselines extensively, which justifies the effectiveness of our model.
	\item ``Correlated-ATT'' performs better than ``Softmax-ATT'', which means to characterize the correlations between user and item would improve the model performance.
	\item There exist some domains, where the promotion is obviously larger than the others. We suppose there exist more clear hierarchical perspectives in these domains. For the example of Music domain, there are many low-level aspects such as singer, writer, composer, volume and speed, based on which, high-level aspects such as genre, style, melody are constructed and analyzed.
\end{itemize}

\subsection{Sensitive to Hyper-Parameters}
In this subsection, in order to verify the effect of hyper-parameters, we leverage the ``Correlated-ATT'' setting for attention mechanism and also the optimal experimental setting that are introduced in Implementation as default. 

\begin{figure}[H]
	\centering
	\includegraphics[width=0.9\linewidth]{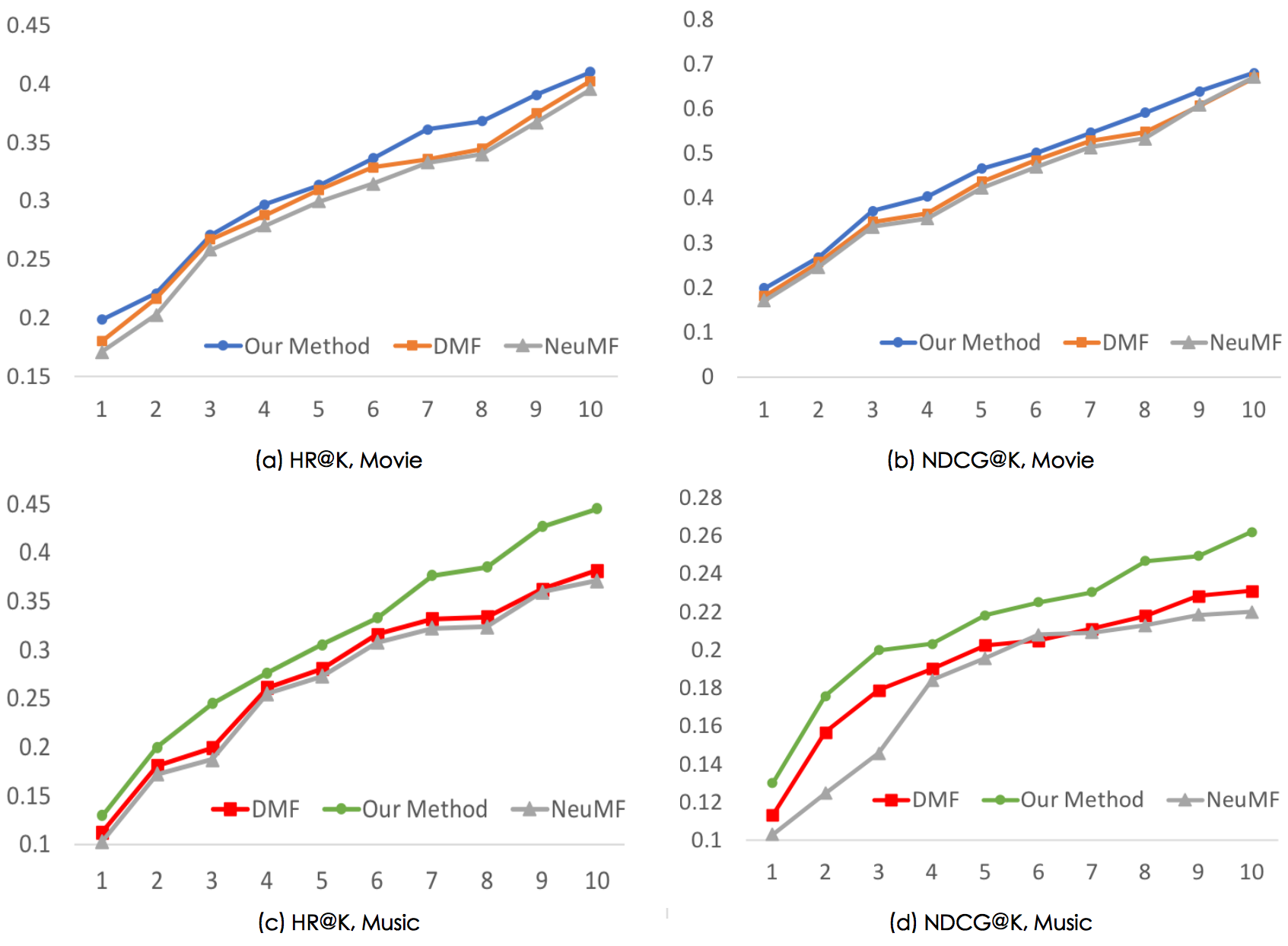}
	\caption{Evaluation of Top-$K$ item recommendation, where $K$ ranges from $1$ to $10$ on the datasets of Movie and Music. The y-axis of (a) and (c) is HR@K, while that of (b) and (d) is NDCG@K. The x-axis of all the sub-figures is the $K$ of top-$K$.}
	\label{fig:varyk}
\end{figure}

\textbf{HR@K \& NDCG@K.}  Fig.\ref{fig:varyk} shows the performance of top-$K$ recommended lists where the ranking position $K$ ranges from $1$ to $10$. As can be concluded, our method demonstrates consistent improvements over other methods across different $K$. For the dataset of Movie, our model outperforms DMF by 0.0239 for HR@K and 0.010 for NDCG@K in average, while for the dataset of Music, our method promotes DMF by 0.0360 for HR@K and 0.0261 for NDCG@K in average. This comparison demonstrates the consistent effectiveness of our methods.

\begin{figure*}
	\centering
	\includegraphics[width=0.9\linewidth]{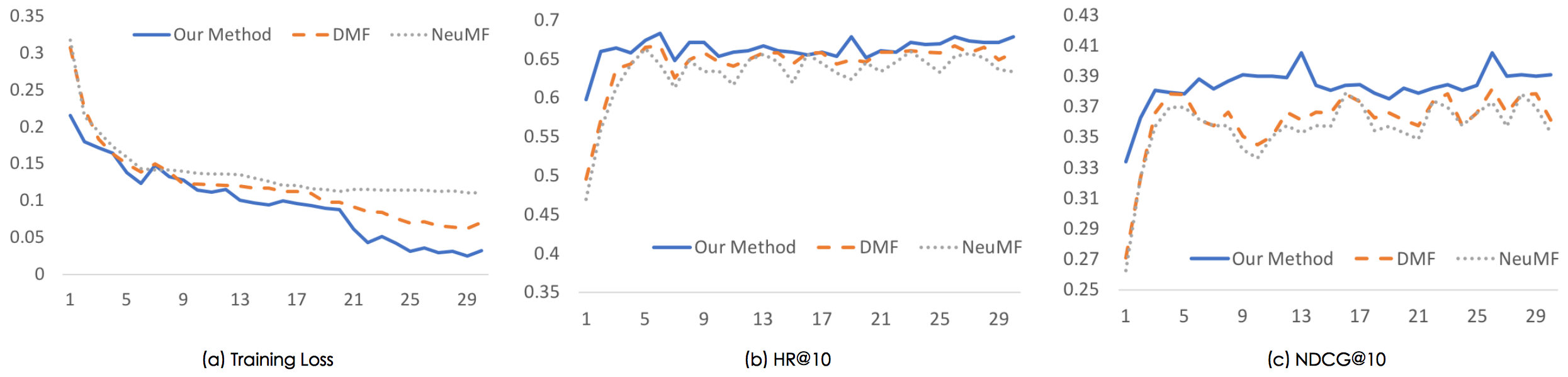}
	\caption{The training loss (averaged over all the training instances), HR@10 and NDCG@10 over iterations on the dataset of Movie.}
	\label{fig:traingingloss}
\end{figure*}

\textbf{Effect of Number of Negative Samples.}  
Argued in the previous section, our method samples negative instances from unobserved data for training. In this experiment, different negative sampling ratios are tested for the performance variance (e.g neg-5 indicates that the negative sampling ratio is 5 or we sample 5 negative instances per positive instance). From the results in Tab.\ref{neg}, we discover that larger negative sample ratio could lead to better performance, while overlarge ratio seems to harm the results. For the example of NDCG on the dataset of Movie, the performance increases before neg-5, while it drops after neg-9. In detail, the optimal negative sample ratio is around 5, which consistently accords to the previous researches, \cite{he2017generating, Xue2017Deep}.

\begin{table}
	\centering
	\renewcommand\arraystretch{1.1}
	\begin{tabular}{ccccccc}
		\hline 
		\hline
		\multirow{2}{*}{Datasets} & \multirow{2}{*}{Metric} & \multicolumn{5}{c}{Negative Sample Ratio} \\
		\cline{3-7} & & 1 & 2 & 5 & 9 & 10 \\
		\hline \multirow{2}{*}{Movie} & NDCG & 0.342 & 0.351 & 0.368 & 0.367 & 0.364 \\
		& HR & 0.615 & 0.633 & 0.642 & 0.646 & 0.645 \\
		\hline \multirow{2}{*}{Music} & NDCG & 0.202 & 0.205 & 0.217 & 0.224 & 0.216 \\
		& HR & 0.341 & 0.345 & 0.360 & 0.372 & 0.359 \\
		\hline \multirow{2}{*}{Baby} & NDCG & 0.168 & 0.169 & 0.170 & 0.173 & 0.182 \\
		& HR & 0.312 & 0.321 & 0.322 & 0.319 & 0.335 \\
		\hline \multirow{2}{*}{Office} & NDCG & 0.235 & 0.235 & 0.254 & 0.240 & 0.244 \\
		& HR & 0.506 & 0.507 & 0.507 & 0.513 & 0.512 \\
		\hline 
		\hline 
	\end{tabular}
	\caption{Results for different negative sampling ratios.}
	\label{neg}
\end{table}

\begin{figure}[H]
	\centering
	\includegraphics[width=0.9\linewidth]{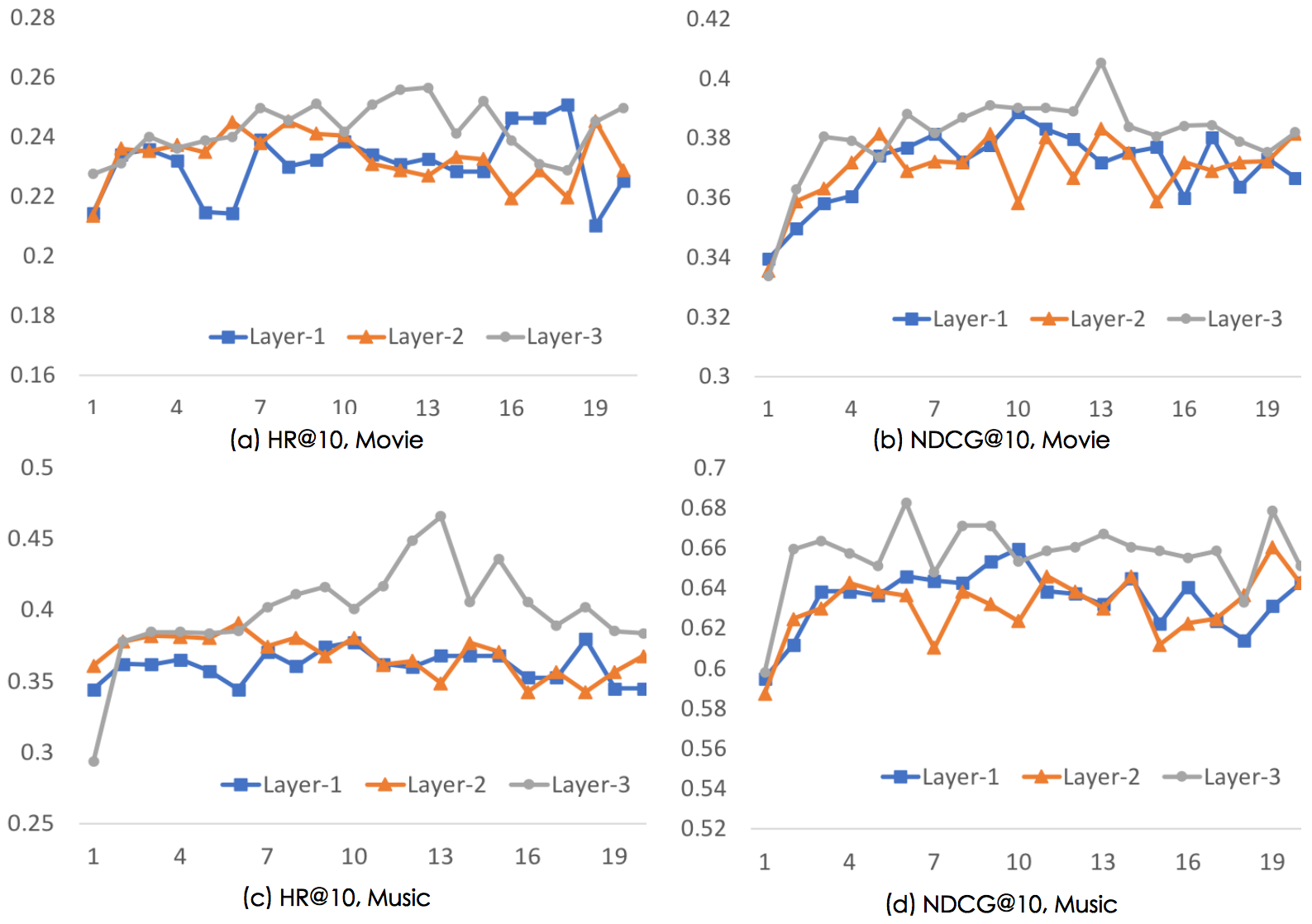}
	\caption{Results for different layers. The y-axis of (a) and (c) is HR@10, while the y-axis of (b) and (d) is NDCG@10. The x-axis is training epoch.}
	\label{fig:layer}
\end{figure}

\textbf{Effect of Number of Layers.} Since we model the hierarchically organized perspectives, the depth or the layer number could be a critical factor in our method. Thus, we conduct experiments to test the effect of depth. Shown in Fig.\ref{fig:layer}, we could conclude that the 3-layer architectures work best among all the present models. Specifically, on the dataset of Movie, the optimal performance of layer-3 outperforms that of layer-2 by 0.021 for HR and 0.019 for NDCG, while on the dataset of Music, the optimal performance of layer-3 improves that of layer-2 by 0.072 for HR and 0.014 for NDCG. Thus, we conjecture deeper models could extract more abstract perspectives, which help to boost the performance. 

\textbf{Effect of Final Latent Dimension.} Besides the negative sample ratio and the number of layers, the final latent dimension is also a sensitive factor, which directly guides the generation of predicted user's preference. We vary the final latent dimension from $8$ to $128$ for the experiments. Demonstrated in Tab.\ref{dim}, we observe that larger final dimension leads to better performance. For the example of Movie dataset, HR increases with latent dimension number. Thus, we suppose larger latent dimension could encode more information into the final results, which could lead to better prediction accuracies. 

\begin{table}
	\centering
	\renewcommand\arraystretch{1.1}
	\begin{tabular}{ccccccc}
		\hline 
		\hline
		\multirow{2}{*}{Datasets} & \multirow{2}{*}{Metric} & \multicolumn{5}{c}{Final Latent Dimension} \\
		\cline{3-7} & & 8 & 16 & 32 & 64 & 128 \\
		\hline \multirow{2}{*}{Movie} & NDCG & 0.392 & 0.395 & 0.400 & 0.390 & 0.410 \\
		& HR & 0.656 & 0.667 & 0.663 & 0.687 & 0.688 \\
		\hline \multirow{2}{*}{Music} & NDCG & 0.241 & 0.246 & 0.248 & 0.250 & 0.262  \\
		& HR & 0.383 & 0.392 & 0.430 & 0.433 &  0.445 \\
		\hline \multirow{2}{*}{Office} & NDCG & 0.263 & 0.248 & 0.273 & 0.276 & 0.262 \\
		& HR & 0.526 & 0.514 & 0.525 & 0.523 & 0.532 \\
		\hline \multirow{2}{*}{Book} & NDCG & 0.476 & 0.480 & 0.480 & 0.488 & 0.484 \\
		& HR & 0.689 & 0.690 & 0.691 & 0.692 & 0.694 \\
		\hline 
		\hline 
	\end{tabular}
	\caption{Results for different final latent factor numbers.}
	\label{dim}
\end{table}

\textbf{Training Loss and Performance.} Fig.\ref{fig:traingingloss} shows the training loss (averaged over all the training instances) and recommendation performance of our method and state-of-the-art baselines of each iteration on the dataset of Movie. Results on the other datasets show the same trend, thus they are omitted for limited pages. From the results, we could draw two observations. First, we could see that with more iterations, the training loss of our method gradually decreases and the recommendation performance is promoted. The most effective updates are in first 10 iterations and more iterations increase the risk of overfitting, which accords to our common knowledge. Second, our method achieves the lower training loss than DMF, which illustrates that our model could fit the data in a better degree. Thus, a better performance over DMF is expected. Overall, the experiments show the effectiveness of our method.

\section{Conclusion}
In this paper, we propose a novel neural architecture for recommendation system. Our model encodes the user and item from multiple hierarchically organized perspectives with attention mechanism and then metrics the abstract representations to predict the user's preference on the item. Extensive experiments on several benchmark datasets demonstrate the effectiveness of our proposed methods. 
We will publish our poster, slides, datasets and codes at \url{https://www.github.com/...}.


\newpage


\bibliography{MPDMF}
\bibliographystyle{aaai}
	
\end{document}